\documentclass[conference]{IEEEtran}
\IEEEoverridecommandlockouts
\usepackage{cite}
\usepackage{amsmath,amssymb,amsfonts,amsthm}
\usepackage{graphicx}
\usepackage[table,xcdraw]{xcolor}
\usepackage{textcomp}
\def\BibTeX{{\rm B\kern-.05em{\sc i\kern-.025em b}\kern-.08em
    T\kern-.1667em\lower.7ex\hbox{E}\kern-.125emX}}

\usepackage{lettrine}
\usepackage{comment}
\usepackage{enumitem}
\usepackage{booktabs}
\usepackage[table]{xcolor}
\usepackage{multirow}
\usepackage{tablefootnote}
\usepackage{bbm}
\usepackage{bm}
\usepackage{chngpage}
\usepackage{textcomp}
\usepackage{hyperref}
\usepackage{url}
\usepackage[normalem]{ulem}
\usepackage{algpseudocode}
\newsavebox{\ieeealgbox}

\usepackage{array}

\newtheorem{theorem}{Theorem}
\newtheorem{proposition}{Proposition}
\newtheorem{corollary}{Corollary}

\newtheorem{definition}{Definition}

\newtheorem*{policy*}{Dynamic NEM}

\newcommand*{\QED}{\hfill\ensuremath{\square}}

 \def\old#1{}

\def\nn{\nonumber}
\def\beq{\begin{equation}}
\def\eeq{\end{equation}}
\def\bea{\begin{eqnarray}}
\def\eea{\end{eqnarray}}
\def\ba{\begin{array}}
\def\ea{\end{array}}

\def\bitem{\begin{itemize}}
\def\eitem{\end{itemize}}
\def\ben{\begin{enumerate}}
\def\een{\end{enumerate}}

\definecolor{bgrd}{rgb}{1,1,1}
\definecolor{gray}{rgb}{0.5,0.5,0.5}
\definecolor{dkr}{rgb}{0.7,0.1,0.2}
\definecolor{dkb}{rgb}{0.1,0.1,0.8}

\def\dbf{{\bf d}}

\def\Lbf{{\bf L}}

\def\Gc{{\cal G}}

\def\Pc{{\cal P}}

\begin{document}

\title{Co-optimizing Behind-The-Meter Resources under Net Metering
}

\author{\IEEEauthorblockN{Ahmed S. Alahmed and Lang Tong (\{asa278, lt35\}@cornell.edu)}
\IEEEauthorblockA{\textit{School of Electrical and Computer Engineering, Cornell University}, Ithaca, USA}
}

\maketitle

\begin{abstract}
We consider the problem of co-optimizing behind-the-meter (BTM) storage and flexible demands with BTM stochastic renewable generation. Under a generalized net energy metering (NEM) policy---NEM X, we show that the optimal co-optimization policy schedules the flexible demands based on a load priority list that defers less prioritized loads to times when the BTM generation is abundant. This gives rise to the notion of a net-zero zone, which we quantify under different distributed energy resources (DER) compositions. We highlight the special case of inflexible demands that results in a storage policy that minimizes the imports and exports from and to the grid. Comparative statics are provided on the optimal co-optimization policy. Simulations using real residential data show the surplus gains of various customers under different DER compositions.
\end{abstract}

\begin{IEEEkeywords}
demand response, distributed energy resources, energy storage, home energy
management, net metering.
\end{IEEEkeywords}

\section{Introduction}\label{sec:intro}
\lettrine{T}{he} falling prices of battery storage and the unremitting reduction of NEM compensation rates for grid exports ushered storage deployment in residential households, especially those coupled with rooftop solar\footnote{According to \cite{Barbose&Elmallah&Gorman:21LBNL}, the total installed BTM storage capacity in 2020 reached 1000 MW, and 80\% of the residential storage capacity is paired with solar.}. The increasing differential between the rates of energy imports and exports under NEM, increases the value of self-consuming the BTM generation, which can be achieved by demand response \cite{Alahmed&Tong:22IEEETSG}, energy storage \cite{Darghouth&Barbose&Mills:19LBNL}, or both \cite{Alahmed&Tong:22arXiv}.

\par Substantial research studied home energy management systems (HEMS) under the existence of DER. The work in \cite{Alahmed&Tong:22arXiv}, however, was the first to propose a linear-complexity and near closed-form characterization to the storage and flexible demand co-optimization, which makes it possible to schedule a large number of flexible demands and storage decisions as functions of the BTM renewable distributed generation (DG). In this work, we expand on the analysis in \cite{Alahmed&Tong:22arXiv} by deriving additional structural properties and special cases that give more insights to the optimal co-optimization policy.

The optimization of BTM storage operation has been extensively studied. Researchers have studied the optimal operation of storage for various objectives including bill minimization (or surplus maximization) \cite{Peddrasa&etal:10TSG,Xu&Tong:17TAC,Guo&etal:13TSG}, wholesale market participation \cite{Gonzalez&etal:08TPS}, and grid services provision \cite{Hao&Wu&Lian&Yang:18TSG}. The literature on BTM residential storage operation largely omitted the situation of dynamically controlling both the household's consumption and storage to maximize the customer's benefit by actively scheduling the resources as functions of the renewable DG. The handful of work done on consumption and storage co-optimization either lacks the analyticity of our proposed solution \cite{Peddrasa&etal:10TSG,Li&Chen&Low:11PESGM,Gonzalez&etal:08TPS}, or co-optimizes storage and
time-of-service of deferrable load \cite{Xu&Tong:17TAC,Guo&etal:13TSG} rather than the quantity of the elastic load, as presented in our work.

This work builds upon the optimal policy derived in \cite{Alahmed&Tong:22arXiv}, which co-optimizes flexible loads and energy storage under the existence of stochastic renewable generation when prosumers face NEM X tariff, by providing insights, interpretations, and special cases to the solution structure. To this end, four results are presented. First, we show that the optimal co-optimization policy generates a load priority list that schedules demands based on the DG profile. Second, we prove that the special case of inflexible demands simplifies the optimal policy to one that minimizes the prosumer's inflows and outflows from and to the grid. Third, we quantify the net-zero zone, where the prosumer is operationally off-the-grid, under different DER compositions. Lastly, we perform comparative statics on the co-optimization policy's optimal decisions

Our simulation results adopts the Californian NEM 3.0 policy\footnote{Also called {\em net billing} as stated in the California PUC \href{https://docs.cpuc.ca.gov/PublishedDocs/Efile/G000/M498/K526/498526033.PDF}{decision}.} to model household's payment, under various customers types, including those with and without storage, DG and flexible demands. The surplus gain and percentage of self-consumed DG of the different customer types are investigated while varying tariff and storage parameters.

\section{Problem Formulation and Optimal Decisions}\label{sec:formulation}
\begin{figure}[htbp]
    \centering
    \includegraphics[scale=0.51]{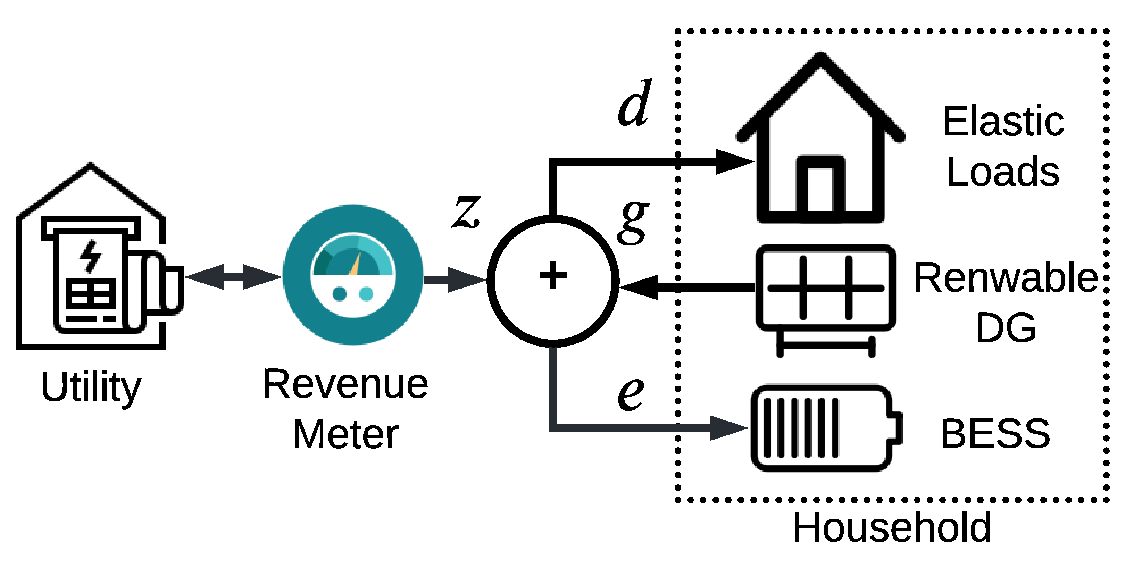}
    \caption{Solar+storage prosumer under NEM. The variables of consumption $d$ and DG output $g$ are real and non-negative, whereas storage output $e$ and net consumption $z$ variables are real.}
    \label{fig:MeteringArchitecture}
\end{figure}

We consider a household with BTM DER, including flexible demands, a renewable DG, and an energy storage, facing an electric utility under NEM (Fig.\ref{fig:MeteringArchitecture}). The surplus-maximizing household co-optimizes its BTM DER, which reveals appealing decision and operational structures that we investigate.

\subsection{BTM DER Models}
For the system model in Fig.\ref{fig:MeteringArchitecture}, we consider the sequential scheduling of consumption $(\dbf_t)$ and storage operation $(e_t)$  over a finite horizon indexed by $t=0,\ldots, T-1$.
\paragraph{Flexible demand}  The household has $K$ devices whose consumption bundle for $t=0,\ldots, T-1$ is denoted by ${\bm d}_t = (d_{t1},\cdots, d_{tK}) \in \mathcal{D}$, with
$$\mathcal{D}:=\{\bm{d}:\underline{\bm{d}}\preceq \bm{d} \preceq \overline{\bm{d}}\} \subseteq \mathcal{R}^K_+,$$ where  $\underline{\bm{d}},\overline{\bm{d}}$ are the consumption bundle's lower and upper limits, respectively. A device $k$ is uncontrollable if $\underline{d}_{k}= \overline{d}_{k}$. The household's {total consumption} is defined as $d_t:= \bm{1}^\top \bm{d}_t$.  

The utility $U_t(\dbf_t)$ of consuming $\dbf_t$ in interval $t$ is assumed to be strictly concave, strictly increasing, continuously differentiable, and additive, i.e.,
\beq \label{eq:U}
U_t(\dbf_t) := \sum_{k=1}^K U_{tk}(d_{tk}), \quad t=0,\ldots, T-1.
\eeq
The marginal utility function is denoted and defined by $\Lbf_t:=\nabla U_t=(L_{t1},\cdots, L_{tK})$.

\paragraph{Renewable}  For $t=0,\ldots, T-1$, the {\em renewable generation} $g_t \in \mathcal{R}_+$ is an exogenous (positive) Markovian random process.

\paragraph{Battery storage}   For $t=0,\ldots, T-1$, the {\em storage control} is denoted by $e_t = [e_t]^+ - [e_t]^- \in [-\underline{e},\bar{e}] $, where $[x]^+:=\max\{0,x\}$ and $[x]^-:=-\min\{0,x\}$ denote the positive and negative part functions for any $x \in \mathcal{R}$, respectively, and
 $\underline{e}$ and $\bar{e}$ are the maximum energy discharging and charging rate constraints, respectively.  The battery is charged when $e_t >0$ and discharged when $e_t<0$.

The charging and discharging efficiencies are denoted by $\tau \in (0,1]$ and $\rho \in (0,1]$, respectively,  which means that charging the storage with $[e_t]^+$ results in an increase of the SoC by $\tau[e_t]^+$, whereas discharging by $[e_t]^-$ reduces the SoC by  $\frac{1}{\rho}[e_t]^-$. The storage SoC is denoted by $s_t \in [\underline{s},\bar{s}]$ with $\underline{s}$ and $\bar{s}$ as the lower and upper capacity limits, respectively. The SoC evolution is driven by $e_t$ as
\begin{equation}\label{eq:SOCevolution}
    s_{t+1} = s_t + \tau [e_t]^+-[e_t]^-/\rho, \quad t=0,\ldots, T-1,
\end{equation}
with $s_0=s$ as the initial SoC, which is assumed to be exogenous and independent of the household's decisions.

\subsection{NEM X Tariff Model}
The customer's payment to the utility under NEM policy is based on the household's {\em net-consumption} $z_t \in \mathcal{R}$ defined by
\begin{equation}\label{eq:NetConsumption}
z_t:=d_t + e_t-g_t, \quad t=0,\ldots, T-1.
\end{equation}
The net billing period can be as short as 5 minutes and as long as a day or a month \cite{Alahmed&Tong:22EIRACM}.  For ease of presentation, we restrict ourselves to the case that the NEM billing period is the same as the household's scheduling period, which allows us to index the billing period also by $t$.

 By adopting the NEM X tariff model introduced in \cite{Alahmed&Tong:22IEEETSG}, we use the NEM X tariff parameter $\pi_t=(\pi^+_t,\pi^-_t, \pi^0_t)$, to compute the customer's payment under NEM X as
\begin{equation}\label{eq:NEMpayment}
    P^{\pi_t}_t(z_t) := \pi^{+}_t [z_t]^+-\pi^{-}_t [z_t]^- +\pi^{0}_t,~~ t=0,\ldots, T-1,
\end{equation}
where $\pi^+_t,\pi^-_t,\pi^0_t \in \mathcal{R}_+$ are the {\em retail (buy) rate}, {\em export (compensation) rate}, and {\em fixed (connection) charge}\footnote{Without loss of generality, we assume zero fixed charges ($\pi^0_t=0, \forall t$), since it does not affect the optimization or solution structure.}, respectively. The prosumer is a \textit{net-consumer} and faces $\pi^+_t$ when $z_t\geq 0$, and a \textit{net-producer} facing $\pi^-_t$ when $z_t<0$.

\par The retail and export rates can be temporally varying (e.g., time-of-use (ToU)) or fixed (e.g., flat pricing).

\subsection{Prosumer Decision Problem}
We formulate the prosumer decision problem as a $T$-stage Markov decision process (MDP).  The state of the MDP in interval $t$ includes battery SoC $s_t$ and renewables $g_t$,  $x_t :=(s_t, g_t) \in \mathcal{X}$,  whose evolution is defined by (\ref{eq:SOCevolution}) and the exogenous Markov process  $(g_t)$. The initial state is denoted by $x_0=(s, g)$.

\par An MDP {\em policy} $\mu := (\mu_0,\ldots,\mu_{T-1})$ is a sequence of decision rules, $x_t \stackrel{\mu_t}{\rightarrow} u_t := (\bm{d}_t,e_t)$, for all $x_t$ and $t$, that specifies consumption and storage operation in each interval. The reward function $r_{t}^{\pi_t}$ consists of {\em prosumer surplus} $S^{\pi_t}_t$ as a stage reward, and {\em storage salvage value} as a terminal reward:
\beq
        r_{t}^{\pi_t}\left(x_t,u_t\right) := \begin{cases} S^{\pi_t}_t(u_t;g_t),&\hspace{-0.2cm} t \in[0, T-1] \\ \gamma (s_{T}-s), &\hspace{-0.2cm} t=T,\end{cases} \label{eq:stageReward},\\
\eeq
where 
\beq\label{eq:Surplus}
S^{\pi_t}_t(u_t;g_t):=U_t(\bm{d}_t)-P^{\pi_t}_t(\bm{1}^\top \bm{d}_t-g_t+e_t),
\eeq
and $\gamma$ is the (marginal) salvage value of stored energy.

 The storage-consumption co-optimization is defined by
 \begin{subequations}\label{eq:optimization}
\begin{align}
   \Pc: \underset{\mu = (\mu_0,\ldots,\mu_{T-1})}{\text { Maximize}} &~~ \mathbb{E}_{\mu}\left\{\gamma (s_{T}-s)+\sum_{t=0}^{T-1}   r_{t}\left(x_t,u_t\right)\right\} \\\text { Subject to~~} & \mbox{for all}~ t=0,\ldots, T-1,\nn\\&
   s_{t+1}=s_t + \tau [e_t]^+-[e_t]^-/\rho \\&
    g_{t+1} \sim F_{g_{t+1}|g_t} \\&
    \underline{s} \leq s_{t} \leq \bar{s}   \label{eq:SoCcapacity}\\&
    0 \leq [e_t]^- \leq \underline{e} \label{eq:e-}\\&
    0 \leq [e_t]^+ \leq \bar{e}   \label{eq:e+}\\&
    \bm{\underline{d}} \preceq \bm{d}_{t} \preceq \bm{\overline{d}} \label{subeq:DeviceConstraint}   \\&
    x_{0}=(s,g) , \end{align}
\end{subequations}
 where $F_{g_{t+1}|g_t}$ is the conditional distribution of $g_{t+1}$ given $g_t$, and the expectation is taken over the exogenous stochastic generation $(g_t)$.

\subsection{Optimal Prosumer Decisions}\label{sec:OptDec}
The solution of the storage-consumption co-optimization in (\ref{eq:optimization}) is provided in \cite{Alahmed&Tong:22arXiv} under two assumptions: (A1) non-binding SoC limits (\ref{eq:SoCcapacity}), and (A2) sandwiched salvage value, where
\beq \label{eq:sandwichedGamma}
\max\{(\pi_t^-)\} \le \tau\gamma \le \gamma/\rho \le \min\{(\pi_t^+)\}.
\eeq
Under A1-A2, the solution has a highly-scalable threshold-based structure that co-schedules the consumption and storage based on the availability and level of the BTM DG. For every $t=0,\ldots, T-1$, the co-optimization policy has six BTM-DG-independent thresholds, ordered as
\beq \label{eq:ThresholdsOrder}
 \Delta_t^- \geq \sigma_t^- \geq \sigma_t^{-o} \geq \sigma_t^{+o}\geq \sigma_t^+ \geq \Delta_t^+,
\eeq
and computed as:
\beq \label{eq:DeltaSigma}
\begin{array}{ll}
\Delta_t^+:= f_t(\pi_t^+)-\underline{e},~~&\Delta_t^-:= f_t(\pi_t^-) + \overline{e},\\
\sigma_t^+:= f_t(\gamma/\rho)- \underline{e}, &\sigma_t^-:= f_t(\tau\gamma) + \overline{e},\\
\sigma_t^{+o}:= f_t(\gamma/\rho), &\sigma_t^{-o}:= f_t(\tau\gamma),\\
\end{array}
\eeq
where $f_{t}$ is the sum of inverse marginal utilities defined as
\bea
f_t(\pi_t) &:=& \sum_{k=1}^K f_{tk}(\pi_t)\label{eq:ft}\\ f_{tk}(\pi_t) &:=&\max\{\underline{d}_k,\min\{L_{tk}^{-1}(\pi_t),\overline{d}_k\}\},\label{eq:ftk}
\eea
and $L_{tk}$ is the marginal utility function (\ref{eq:U}) of device $k$ in interval $t$ and $L_{tk}^{-1}$ its inverse. 

For every $t=0,\ldots, T-1$, the {\em optimal consumption} $d^\ast_{tk}(g_t)$ of every device $k$, {\em storage operation} $e^\ast_{t}(g_t)$, {\em net-consumption} $z^\ast_{t}(g_t)$, and the resulting payment $P^{\ast,\pi_t}_t(g_t)$ are monotonic in $g_t$, and their structures are summarized in Table \ref{tab:OptSchedule} (Also depicted in Fig.\ref{fig:ActivePassiveSDG}). The household operates in 1) the net consumption zone $(+)$ if $g_t\leq \Delta^+_t$, 2) the net production zone $(-)$ if $g_t\geq \Delta^-_t$, and 3) the net zero zone $(0)$ if $g_t\in  [\Delta^+_t,\Delta^-_t]$, under which the household is off-the-grid. 

\begin{table*}[ht]
\centering
\caption{Summary of the optimal co-optimization policy.}
\label{tab:OptSchedule}
\resizebox{0.95\textwidth}{!}{%
\begin{tabular}{@{}c
>{\columncolor[HTML]{FFF1F0}}c 
>{\columncolor[HTML]{F9F9F9}}c 
>{\columncolor[HTML]{F9F9F9}}c 
>{\columncolor[HTML]{F9F9F9}}c 
>{\columncolor[HTML]{F9F9F9}}c 
>{\columncolor[HTML]{F9F9F9}}c 
>{\columncolor[HTML]{F4FFF4}}c @{}}
\toprule \midrule
Zone & $(+)$ & \multicolumn{5}{c}{\cellcolor[HTML]{F9F9F9}$(0)$} & $(-)$ \\ \midrule
$g_t$ & $g_t < \Delta^+$ & $g_t \in (\Delta^+_t,\sigma^+_t]$ & $g_t \in (\sigma^+_t,\sigma^{+o}_t]$ & $g_t \in (\sigma^{+o}_t, \sigma^{-o}_t]$ & $g_t \in (\sigma^{-o}_t, \sigma^{-}_t]$ & $g_t \in (\sigma^{-}_t, \Delta^-_t]$ & $g_t > \Delta^-_t$ \\
$d^\ast_{tk}(g_t)$ & $f_{tk}(\pi^+_t)$ & $f_{tk}(f^{-1}_t(g_t+\underline{e}))$ & $f_{tk}(f^{-1}_t(\gamma/\rho))$ & $f_{tk}(f^{-1}_t(g_t))$ & $f_{tk}(f^{-1}_t(\gamma \tau))$ & $f_{tk}(f^{-1}_t(g_t-\overline{e}))$ & $f_{tk}(\pi^-_t)$ \\
$e^\ast_t(g_t)$ & $-\underline{e}$ & $-\underline{e}$ & $g_t-\sigma^{+o}_t$ & $0$ & $g_t-\sigma^{-o}_t$ & $\overline{e}$ & $\overline{e}$ \\
$z_t^\ast(g_t)$ & $z_t^\ast>0$ & $z_t^\ast=0$ & $z_t^\ast=0$ & $z_t^\ast=0$ & $z_t^\ast=0$ & $z_t^\ast=0$ & $z_t^\ast<0$ \\
$P^{\ast,\pi_t}_t(g_t)$ & $\pi^+_t z^\ast_t$ & $0$ & $0$ & $0$ & $0$ & $0$ & $\pi^-_t z^\ast_t$ \\ \midrule \bottomrule
\end{tabular}%
}
\end{table*}

\section{Solution Properties and Special Cases}\label{sec:Models}
We discuss some properties and special cases of the optimal co-optimization policy in Sec.\ref{sec:OptDec}. Sec.\ref{sec:LPRR} shows the load priority ranking structure of the solution. Sec.\ref{sec:PassiveSDG} considers the special case of solving (\ref{eq:optimization}) under passive\footnote{We use  \textit{passive} and \textit{active} to refer to customers with DG-inelastic and DG-elastic demands, respectively \cite{Alahmed&Tong:22EIRACM}.} demands. Sec.\ref{sec:NZquantification}, quantifies the net zero zone, and show that the more flexible resources the prosumer has, the larger its net zero zone. Lastly, Sec.\ref{sec:ComparativeStatics}, provides comparative statics analysis on the solution structure.

\subsection{Load Priority Ranking Rule}\label{sec:LPRR}
The optimal consumption schedule reveals important microeconomics interpretations based on marginal utilities of devices and the rates $\pi^+_t, \pi^-_t, \gamma/\rho$ and $\tau \gamma$. Devices with higher marginal utilities are prioritized in the net-consumption zone ($z^\ast_{t}(g_t)>0$); when the DG output is low. Less important devices (lower marginal utilities) are exercised only when the DG output is high. Proposition \ref{prop:loadranking} formalizes the conditions for device consumptions in each net-consumption zone.

\begin{proposition}[Load priority ranking rule] \label{prop:loadranking}
Under A1-A2, and assuming w.l.o.g that $\underline{d}_k=0, \forall k$, the scheduling of every device $k$ and $t=0,\ldots, T-1$ in any of the three consumption zones, depends on its marginal utility $L_{tk}(\cdot)$. If
\begin{enumerate}
    \item $L_{tk}(0) > \pi^+_t $, device $k$ is consumed in all zones.
    \item $\pi^+_t\geq L_{tk}(0)>  \gamma/\rho$, device $k$ is consumed in all zones, except the net-consumption zone.
   \item $\gamma/\rho \geq L_{tk}(0) > \tau \gamma$, device $k$ is consumed only if $g_t > \sigma^{+o}_t$.
  \item $\tau \gamma \geq L_{tk}(0) > \pi^-_t$, device $k$ is consumed only if $g_t > \sigma^{-}_t$.
   \item $\pi^-_t > L_{tk}(0) $, device $k$ is never consumed. \qed
\end{enumerate}
\end{proposition}

\begin{figure}[htbp]
        \centering
        \includegraphics[scale=0.4]{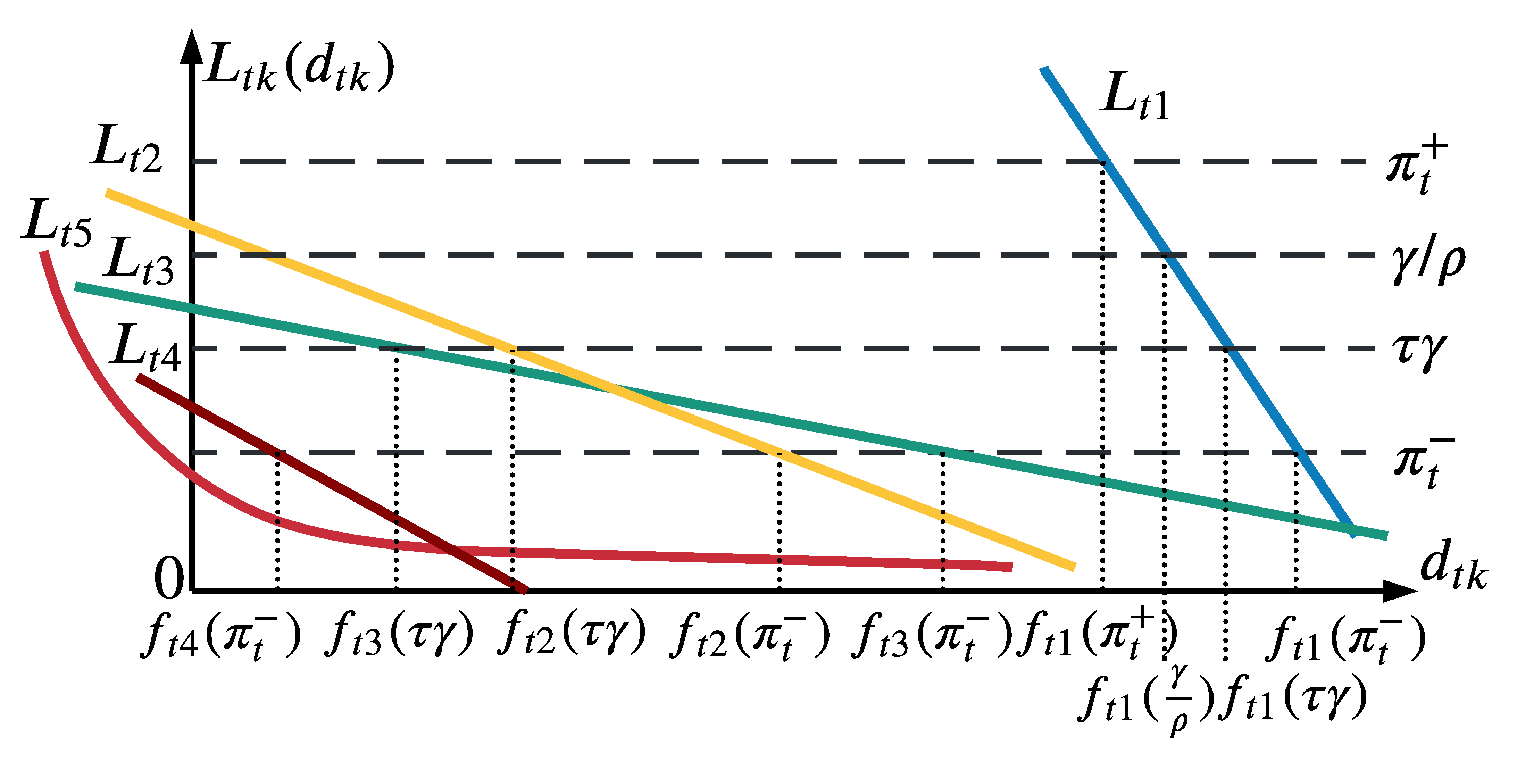}
        \caption{Consumption allocation to devices based on marginal utilities.}
        \label{fig:LoadPriority}
    \end{figure}

To illustrate Proposition \ref{prop:loadranking}, marginal utilities of five devices corresponding to the 5 cases in Proposition \ref{prop:loadranking} are shown in Fig.\ref{fig:LoadPriority}. Note that since the marginal utility of device 1 at zero consumption $L_{t1}(0)$ is greater than $\pi^+_t$, the device was consumed in all zones because the non-increasing marginal utility intersected all of the four price lines, granting a positive consumption. On the other hand,  for device 55, since ($L_{t5}(0)\leq \pi^-_t$), the device was not consumed in any zone. Devices 2, 3 and 4 do not consume in the net-consumption zone since $L_{t2}(0),L_{t3}(0), L_{t4}(0)<\pi^+_t$, however, they start consuming from the smallest point under which their marginal utilities intersect with the price.

\subsection{Passive Prosumer}\label{sec:PassiveSDG}
For every $t=0,\ldots, T-1$, the {\em passive SDG}\footnote{We refer to prosumers with standalone DG as \textit{DG prosumers}, and prosumers with storage and DG as \textit{SDG prosumers}.} prosumer schedules the consumption as if the household faces only $\pi^+_t$, and therefore consumes $f_t(\pi^+_t)$ for any DG output. It turns out that the the optimal policy of a passive SDG prosumer is to minimize the absolute value of its net-consumption $z_t$.

\begin{figure}[htbp]
    \centering
    \includegraphics[scale=0.35]{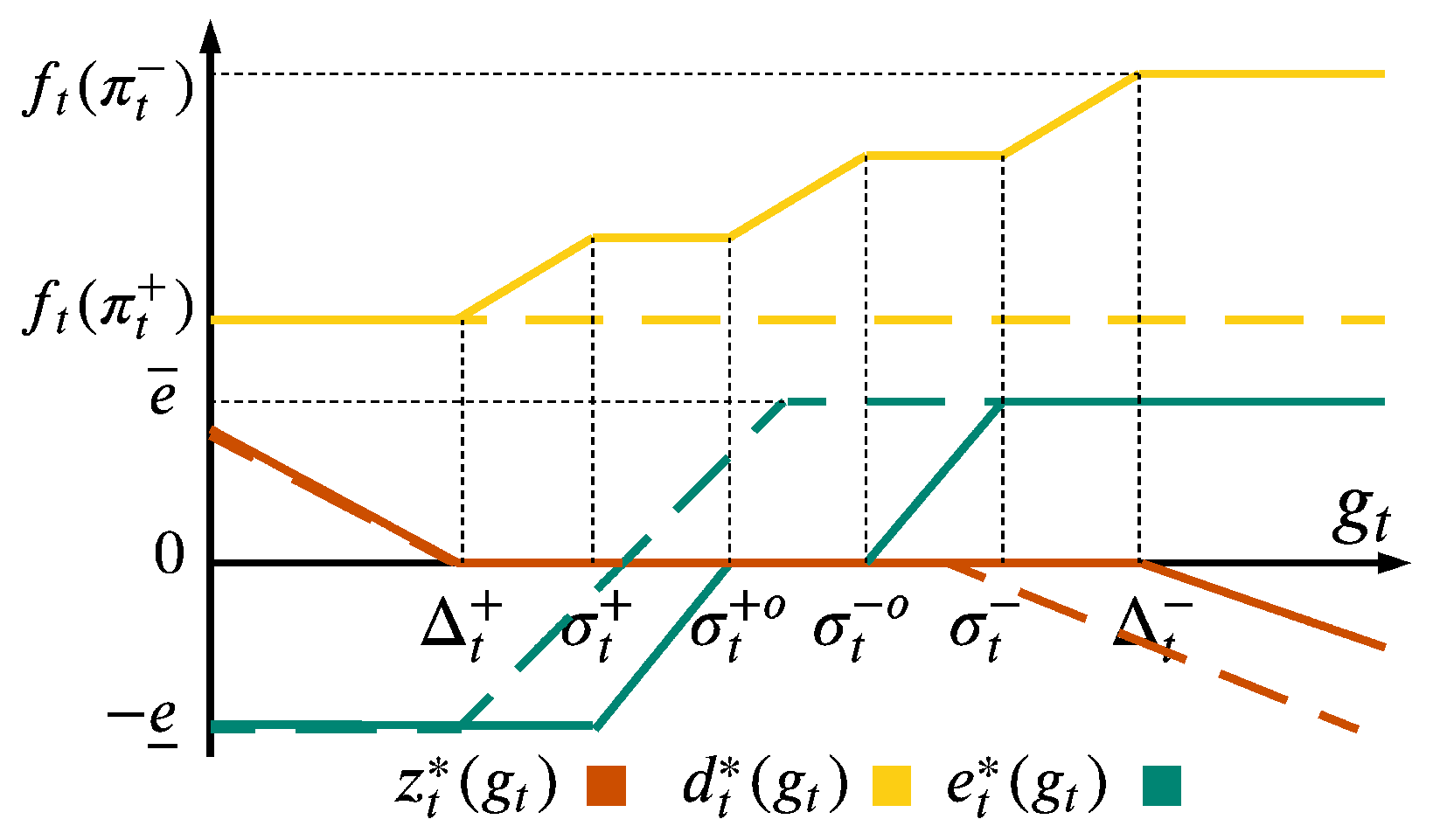}
    \caption{Active (solid) and passive (dashed) SDG prosumers decisions.}
    \label{fig:ActivePassiveSDG}
\end{figure}

\begin{proposition}[Optimal policy under DG-passive demands]\label{prop:ModulusMinimize}
Under A1-A2 and for any DG-passive consumption bundle $\hat{\bm{d}}_t \in \mathcal{D}$, the optimal storage operation is to discharge/charge as much as possible to minimize the absolute value of net consumption:
\begin{align}\label{eq:PassiveOptimization}
\mu^\ast_t \in \underset{e_t \in \{-\underline{e},\bar{e}\}}{\text{arg min}} &~~ |z_t|.
\end{align}
for every $t=0,\ldots, T-1$,
\qed
\end{proposition}
The passive SDG prosumer optimal policy is intuitive (Fig.\ref{fig:ActivePassiveSDG}). The storage exercises a {\em balancing control} that tries to null the renewable-adjusted consumption $\Tilde{d}_t:= \bm{1}^\top \hat{\bm{d}}_t -g_t$ \cite{Harshah&Dahleh:15TPS}. For the given fixed total consumption $f_t(\pi^+_t)$, the battery's stored energy is used to a) reduce net consumption in case $\Tilde{d}_t<0$, b) increase net consumption (production) in case $\Tilde{d}_t>0$, and c) maintain net-zero as much as possible. Under (a), the prosumer gains at the rate of $\pi^+$ and losses at the rate of $\gamma/\rho$, whereas under (b), the prosumer gains at the rate of $\tau \gamma$ and losses at the rate of $\pi^-$. Proposition \ref{prop:ModulusMinimize} also implies that the ratio of self-consumption over the scheduling period $SC \in [0,1]$, defined by 
\begin{equation}\label{eq:SelfConsumption}
    SC(z_t):= 1+\sum_{t=0}^{T-1}\frac{[z_t]^-}{g_t}, \text{ for } \sum_{t=0}^{T-1} g_t>0, 
\end{equation}
is maximized.

\subsection{Net Zero Zone Quantification}\label{sec:NZquantification}
The {\em passive SDG} prosumer case shows the propensity of prosumers to achieve net-zero. We hence define and quantify the net-zero zone length under different DER compositions.
\begin{definition}[Net-zero zone length]\label{def:NetZeroLength}
For $t=0,\ldots,T-1$, and for any optimal policy $\bm{\mu}^\ast$ the net-zero zone length is given by
\begin{equation}\label{eq:NetZeroLength}
   |\Gc_t^{\bm{\mu}^\ast}| := \underset{g_t \in \Gc_t^{\bm{\mu}^\ast}}{\text { Maximize  }} g_t - \underset{g_t \in \Gc_t^{\bm{\mu}^\ast}}{\text { Minimize  }} g_t
\end{equation}
where $\Gc_t^{\bm{\mu}^\ast} = \{g_t \in \mathcal{R}_+: z^{\bm{\mu}^\ast}_t(g_t)=0 \}$ is a convex set.
\end{definition}
Definition \ref{def:NetZeroLength} is used in Corollary \ref{prop:ZeroZoneLengthComparison} to show that the co-optimization policy's net-zero zone length is the sum of the net-zero zone lengths of {\em DG active} \cite{Alahmed&Tong:22IEEETSG} and {\em SDG passive} prosumers (Sec.\ref{sec:PassiveSDG}).
\begin{corollary}[Net-zero zone length quantification]\label{prop:ZeroZoneLengthComparison}
Under A1-A2, and for every $t=0,\ldots,T-1$, the net-zero zone lengths of optimal: 1) passive DG $|\Gc_t^{\bm{\mu}^\ast_1}|$, 2) active DG $|\Gc_t^{\bm{\mu}^\ast_2}|$, 3) passive SDG $|\Gc_t^{\bm{\mu}^\ast_3}|$ and 4) active SDG $|\Gc_t^{\bm{\mu}^\ast_4}|$ prosumers are ordered as
\begin{equation}\label{eq:NZranking1}
    |\Gc_t^{\bm{\mu}^\ast_4}| \geq |\Gc_t^{\bm{\mu}^\ast_3}| \geq |\Gc_t^{\bm{\mu}^\ast_2}| \geq |\Gc_t^{\bm{\mu}^\ast_1}|,
\end{equation}
if $f_t(\pi_t^-)-f_t(\pi_t^+)\leq \overline{e}+ \underline{e}$, and ordered as
\begin{equation}\label{eq:NZranking2}
    |\Gc_t^{\bm{\mu}^\ast_4}| \geq |\Gc_t^{\bm{\mu}^\ast_2}| \geq |\Gc_t^{\bm{\mu}^\ast_3}| \geq |\Gc_t^{\bm{\mu}^\ast_1}|,
\end{equation}
if $f_t(\pi_t^-)-f_t(\pi_t^+) > \overline{e}+ \underline{e}$. 
\par It also holds that
\begin{equation}\label{eq:NZranking3}
    |\Gc_t^{\bm{\mu}^\ast_4}| = |\Gc_t^{\bm{\mu}^\ast_2}|+|\Gc_t^{\bm{\mu}^\ast_3}|.
\end{equation}
\qed
\end{corollary}
The corollary shows that the more flexible resources (active loads and storage) an optimal prosumer has, the longer its net-zero zone length, as in $|\Gc_t^{\bm{\mu}^\ast_4}|$. The length of net-zero zone is compromised if one or both of the decision variables are dropped, i.e., $\bm{d}$ in $|\Gc_t^{\bm{\mu}^\ast_3}|$, $e$ in $|\Gc_t^{\bm{\mu}^\ast_2}|$ and both $\bm{d},e$ in $|\Gc_t^{\bm{\mu}^\ast_1}|$.

Shrinking the net-production and net-consumption zones lengths, while increasing $ |\Gc_t^{\bm{\mu}^\ast}|$ has crucial economical and technical implications. As $|\Gc_t^{\bm{\mu}^\ast}|$ increases, the customer's bill becomes more immune to $\pi^-$ reductions, because $SC$ in (\ref{eq:SelfConsumption}) becomes higher. For grid operators and utilities, higher $|\Gc_t^{\bm{\mu}^\ast}|$ achieves operational benefits such as reducing reverse power flows, which improves network's reliability \cite{RATNAM:15AE}. This, however, may come at the cost of higher grid defection rates \cite{Borenstein_canNetMetering}, requiring utilities to reshape their business model.

\subsection{Comparative Statics}\label{sec:ComparativeStatics}
Here we offer a comparative statics analysis on the optimal policy to investigate how the parameters and variables influence the solution structure in each net-consumption zone. Theorem \ref{thm:ComparativeStatics} in the appendix formalizes the effect of changing exogenous parameters on the endogenous quantities of consumption, storage operation, payment, and surplus. Table \ref{tab:ComparrativeStatics} summarizes the comparative static analysis by considering $\epsilon$-increases of the exogenous parameters and examining changes of endogenous quantities at interior points of the three zones. 

\par Table \ref{tab:ComparrativeStatics} shows that the increase in the renewable output enables increasing the consumption and the storage output, which results in a decreasing payment and increasing surplus. This is because, under NEM X, self-consumption of the renewable output is valued more than exporting it back. 

\par Varying the NEM X tariff parameters $\pi=(\pi^+,\pi^-,\pi^0)$, as shown in Table \ref{tab:ComparrativeStatics} have direct implication on $S^{\ast,\pi_t}_t(g_t)$. Increasing $\pi^+_t$ negatively affects $S^{\ast,\pi_t}_t(g_t)$, since the household consumption will be reduced. However, increasing the export rate positively effect $S^{\ast,\pi_t}_t(g_t)$ as the payment to the utility reduces and the consumption increases. Increasing $\pi^0_t$ reduces $S^{\ast,\pi_t}_t(g_t)$, because the payment increases. Interestingly, the storage output is independent of the NEM X tariff parameters.

\par The salvage value rate affects the optimal consumption, storage operation, and prosumer surplus only in the net-zero zone, where the consumption and surplus reduce as $\gamma$ increases, and the storage output increases as $\gamma$ increases. 

\begin{table}[htbp]
\centering
\caption{Comparative Static Analysis.}
\label{tab:ComparrativeStatics}
\resizebox{0.85\columnwidth}{!}{%
\begin{tabular}{|c|c|c|c|c|c|c|}
 \hline
Quantity & Zone & $g_t \boldsymbol{\uparrow} $ & $\pi^+_t  \boldsymbol{\uparrow}$ & $\pi^-_t \boldsymbol{\uparrow}$ & $\gamma \boldsymbol{\uparrow}$ & $\pi^0_t \boldsymbol{\uparrow}$ \\ \hline
\multirow{3}{*}{$d^\ast_{tk}(g_t)$} & $+$ & --- & $\boldsymbol{\downarrow}$ & --- & --- &  ---\\ \cline{2-7} 
 & $-$ & --- & --- & $\boldsymbol{\downarrow}$ & --- & --- \\ \cline{2-7} 
 & $0$ & $\;\;\boldsymbol{\uparrow}$ & --- & --- & $\boldsymbol{\downarrow}$ & --- \\ \hline
\multirow{3}{*}{$e^\ast_{t}(g_t)$} & $+$ & --- & --- & --- & --- & --- \\ \cline{2-7} 
 & $-$ & --- & --- & --- & --- & --- \\ \cline{2-7} 
 & $0$ & $\;\;\boldsymbol{\uparrow}$ & --- & --- & $\;\;\boldsymbol{\uparrow}$ & ---  \\ \hline
\multirow{3}{*}{$P^{\ast,\pi_t}_t(g_t)$} & $+$ & $\boldsymbol{\downarrow}$ & $\times$ & --- & --- & $\;\;\boldsymbol{\uparrow}$ \\ \cline{2-7} 
 & $-$ & $\boldsymbol{\downarrow}$ & --- & $\boldsymbol{\downarrow}$ & --- & $\;\;\boldsymbol{\uparrow}$ \\ \cline{2-7} 
 & $0$ & --- & --- & --- & --- &  $\;\;\boldsymbol{\uparrow}$ \\ \hline
\multirow{3}{*}{$S^{\ast,\pi_t}_t(g_t)$} & $+$ & $\;\;\boldsymbol{\uparrow}$ & $\boldsymbol{\downarrow}$ & ---  & --- & $\boldsymbol{\downarrow}$  \\ \cline{2-7} 
 & $-$ & $\;\;\boldsymbol{\uparrow}$ & --- & $\;\;  \boldsymbol{\uparrow}$ & --- &  $\boldsymbol{\downarrow}$\\ \cline{2-7} 
 & $0$ & $\;\;\boldsymbol{\uparrow}$  & --- & --- & $\boldsymbol{\downarrow}$ & $\boldsymbol{\downarrow}$  \\  \hline 
\end{tabular}%
}
\footnotesize{ \begin{flushleft} \vspace{0.0cm}
\hspace{0.25cm} $\boldsymbol{\uparrow}:$ increasing $\;$ $\boldsymbol{\downarrow}\; :$ decreasing $\;$ --- : unchanged $\;$ $\times:$ indeterminant
 \end{flushleft}}
\end{table}

\section{Numerical Results}\label{sec:num}
We consider a household receiving service under a NEM policy. Five prosumer types are studied: 1) {\em consumers}: customers without BTM DER, 2) {\em active DG prosumers}: customers who optimize their consumption based on available DG \cite{Alahmed&Tong:22IEEETSG}, 3) {\em active SDG prosumers}: customers who co-optimize storage and consumption as in Sec.\ref{sec:OptDec}, 4) {\em passive DG} and 5) {\em passive SDG prosumers}: customers who do not optimize consumption based on available DG; DG is only used to reduce cost.  {\em Passive prosumers} consume as if they are {\em consumers}.

To model household's consumption and renewable generation, we used the Smart project data set\footnote{The data repository can be accessed at \href{https://traces.cs.umass.edu/index.php/Smart/Smart}{Smart Data Set}. We used home D.}, which has a 1-minute granularity of aggregated and individual home circuits collected over the year of 2016. We restricted our simulation to only three months of 2016 (June-August).

The consumption preferences are captured by the following, widely-adopted, quadratic concave utility function
\begin{equation}
    U_{tk}(d_{tk})=\begin{cases}
        \alpha_{tk} d_{tk}-\frac{\beta_{tk}}{2}d_{tk}^2, & 0\leq d_{tk}< \frac{\alpha_{tk}}{\beta_{tk}}\\
        \frac{\alpha_{tk}^2}{2\beta_{tk}}, & d_{tk}\geq \frac{\alpha_{tk}}{\beta_{tk}},
    \end{cases}
\end{equation}
where $\alpha_{tk}$ and $\beta_{tk}$ are some utility parameters that are learned using historical consumption and price data by positing an elasticity of demand\footnote{The long-run price elasticity of electricity demand used was -0.21 \cite{ASADINEJAD_Elasticity:18EPSR}.} as in \cite{Alahmed&Tong:22EIRACM}. 

The battery charge and discharge efficiencies $\tau, \rho$ were assumed to be 0.95. Similar to Tesla Powerwall\footnote{The specifications of Tesla Powerwall can be found at \href{https://www.tesla.com/sites/default/files/pdfs/powerwall/Powerwall\%202_AC_Datasheet_en_AU.pdf}{Tesla Powerwall}.} 2, the storage capacity was set to 13.5 kWh. The salvage value rate $\gamma$ was chosen so that A2 in (\ref{eq:sandwichedGamma}) holds.

\par The household faces the Californian NEM 3.0 tariff, which has a ToU-based retail rate $\pi^+$, a dynamic avoided-cost-based export rate $\pi^-$, and a fixed charge of $\pi^0=\$15$/month. For $\pi^+$, we adopted PG\&E 2022 summer E-TOU-B rate schedule, which has peak and off-peak rates of $\pi^+_h=\$0.49$/kWh and $\pi^+_l=\$0.37$/kWh, respectively, and a 16--21 peak period. For $\pi^-$, the 2022 average avoided cost rates, developed by E3 Inc. avoided cost calculator (ACC), were used\footnote{The ACC rates can be accessed at \href{https://www.ethree.com/public_proceedings/energy-efficiency-calculator/}{E3 ACC}.}.

\subsection{Surplus Gain}
We compared the average daily surplus gain achieved by the five prosumer types, using {\em consumers} as the benchmark. Table \ref{tab:SurGain} shows the average percentage gain in daily surplus over that achieved by a {\em consumer} under one-minute and one-hour netting frequencies and three different storage charge/discharge rates $\overline{e}=\underline{e} \in \{0.5,0.75,1\}$. 

Four key observations are in order. First, {\em active SDG} customers achieved the highest surplus gain of all cases. Second, increasing the netting frequency from hourly to minutely basis always resulted in lower surplus gains, as customers became more vulnerable to the lower export rate. Third, the value of being {\em active} were significant, with 8\% surplus gain increase for both {\em DG} and {\em SDG} customers. Lastly, increasing storage rate by 0.25kW for both {\em passive} and {\em active} customers resulted in a roughly 4\% surplus gain.

\begin{table}[htbp]
\centering
\caption{Surplus gain over consumers (\%).}
\label{tab:SurGain}
\resizebox{0.65\columnwidth}{!}{%
\begin{tabular}{@{}ccccc@{}}
\arrayrulecolor{black}\toprule \midrule
DER & Customer & $\overline{e}=\underline{e}$ (kW) & 1-min & 1-hour \\ \midrule
-- & Consumer & 0 & 0 & 0 \\\arrayrulecolor{black!30}\midrule
\parbox[t]{2mm}{\multirow{2}{*}{\rotatebox[origin=c]{90}{DG}}} & Passive & 0 & 69.27 & 70.82 \\
 & Active & 0 & 77.48 & 79.21 \\ \arrayrulecolor{black!30}\midrule
\parbox[t]{2mm}{\multirow{6}{*}{\rotatebox[origin=c]{90}{DG + Storage}}} & \multirow{3}{*}{Passive} & 0.5 & 81.27 & 82.98 \\
 &  & 0.75 & 86.13 & 87.90 \\
 &  & 1 & 90.52 & 92.33 \\ \arrayrulecolor{black!30}\cmidrule(l){2-5}
 & \multirow{3}{*}{Active} & 0.5 & 89.44 & 91.23 \\
 &  & 0.75 & 94.19 & 96.02 \\
 &  & 1 & 98.32 & 100.25 \\ \arrayrulecolor{black}  \midrule \bottomrule
\end{tabular}%
}
\end{table}

\subsection{DG Self-Consumption}
Table \ref{tab:SelfCons} shows the self-consumption (computed as in (\ref{eq:SelfConsumption})) percentage of the studied customer types under a one-minute and one-hour netting frequencies and three different storage charge/discharge rates $\overline{e}=\underline{e} \in \{0.5,0.75,1\}$.

Broadly speaking, customers who achieved high surplus gains in Table \ref{tab:SurGain} managed to achieve high self-consumption percentages (Table \ref{tab:SelfCons}). This is, however, not always the case, as {\em active DG} prosumers achieved higher self-consumption but lower surplus gain compared to {\em passive SDG} prosumers with 0.5kW storage charge/discharge rates. The reason is that, although {\em active DG} prosumers more effectively reduced energy exports, they under-performed in reducing energy imports compared to {\em SDG} prosumers, which is more costly. At 0.75kW and 1kW storage rates, {\em passive SDG} prosumers had both higher surplus gains and higher self-consumption.

Table \ref{tab:SelfCons} shows that increasing the netting frequency decreased the self-consumption percentage, as customers had a shorter banking period for loads to consume the DG output. For both netting frequencies, installing a DG without actively scheduling the consumption based on the available DG, resulted in exporting back more than 57\% of the DG. Actively scheduling the consumption based on the available DG, as proposed \cite{Alahmed&Tong:22IEEETSG}, increased self-consumption to over 55\%. When the prosumer installed storage in addition to the DG, self-consumption increased to more than 61\% when the customer was {\em passive}, and to over 74\% when the prosumer was {\em active}.

\begin{table}[htbp]
\centering
\caption{DG self-consumption (\%).}
\label{tab:SelfCons}
\resizebox{0.65\columnwidth}{!}{%
\begin{tabular}{@{}ccccc@{}}
\arrayrulecolor{black} \midrule \toprule
DER & Customer & $\overline{e}=\underline{e}$ (kW) & 1-min & 1-hour \\ \midrule
-- & Consumer & 0 & -- & -- \\\arrayrulecolor{black!30}\midrule
\parbox[t]{2mm}{\multirow{2}{*}{\rotatebox[origin=c]{90}{DG}}} & Passive & 0 & 41.02 & 42.25 \\
 & Active & 0 & 55.22 & 56.68 \\ \arrayrulecolor{black!30}\midrule
\parbox[t]{2mm}{\multirow{6}{*}{\rotatebox[origin=c]{90}{DG + Storage}}} & \multirow{3}{*}{Passive} & 0.5 & 52.22 & 53.32 \\
 &  & 0.75 & 57.18 & 58.19 \\
 &  & 1 & 61.64 & 62.67 \\ \arrayrulecolor{black!30}\cmidrule(l){2-5}
 & \multirow{3}{*}{Active} & 0.5 & 66.24 & 67.40 \\
 &  & 0.75 & 70.87 & 71.97 \\
 &  & 1 & 74.79 & 76.00 \\ \arrayrulecolor{black} \midrule \bottomrule
\end{tabular}%
}
\end{table}

\subsection{Value of Storage}
\begin{figure}[htbp]
    \centering
    \includegraphics[scale=0.325]{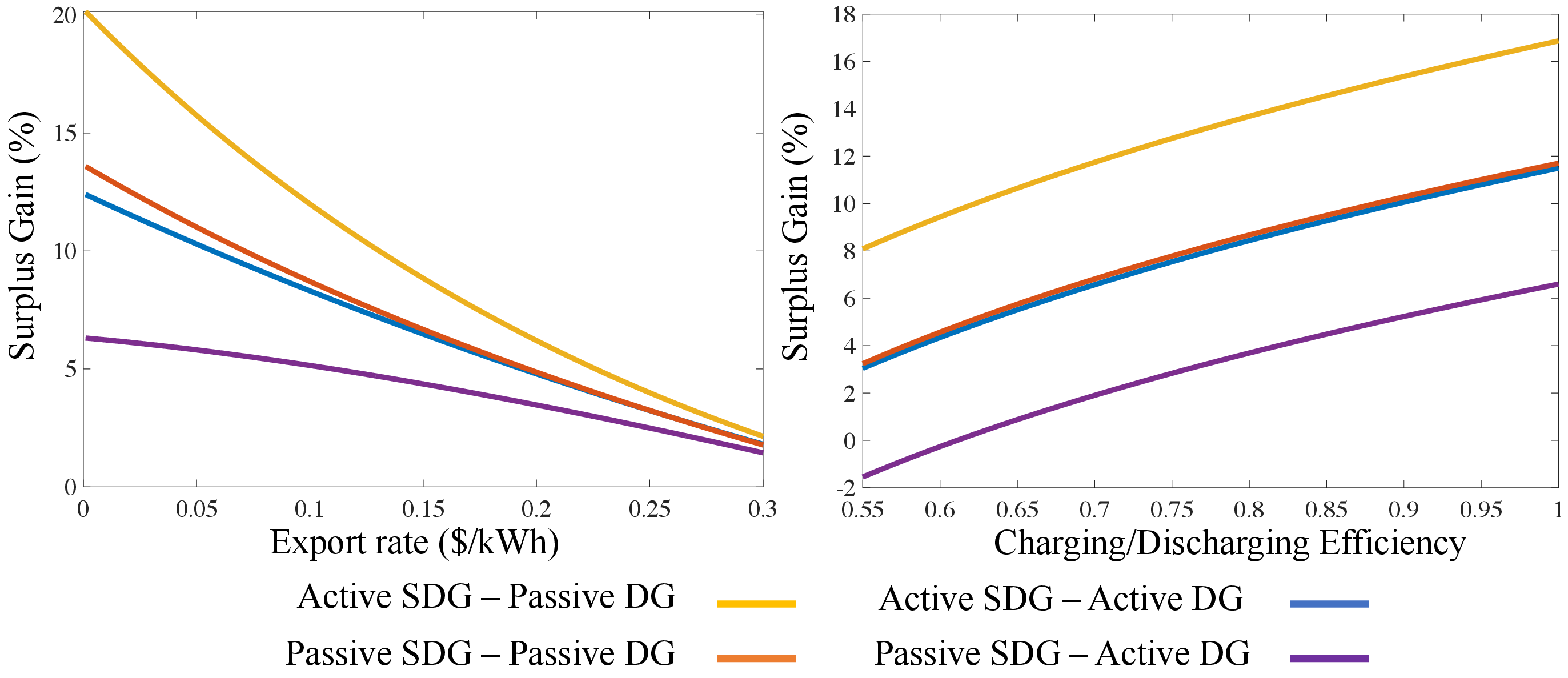}
    \caption{Surplus gain over active/passive DG prosumers ($\overline{e}=\underline{e}=0.75$kW).}
    \label{fig:VoSplot}
\end{figure}

Fig.\ref{fig:VoSplot}, shows surplus gains of {\em active} and {\em passive SDG} prosumers compared to {\em active} and {\em passive DG} prosumers\footnote{This is also called {\em value of storage}.} under varying export rates (left) and storage efficiencies (right). 

\par The left plot shows that the value of storage (VoS) and value of demand response (VDR) increased as the differential between the retail and export rates enlarged. The gain increase was higher in the {\em active SDG -- passive DG} case (yellow), because the curve augmented both VDR and VoS. The surplus gain increased in the {\em passive SDG -- passive DG} (orange) and {\em active SDG -- active DG} (blue) cases, as the $\pi^-$ decreased was primarily due to VoS, which has higher value when locally absorbing $g$ becomes more valuable. The {\em passive SDG -- active DG} surplus gain as $\pi^-$ decreased, compares the effectivity of VoS alone and VDR alone in reducing exported generation.

\par The right plot shows that VoS increased as the storage charging/discharging efficiencies increased. Interestingly, the {\em passive SDG -- active DG} curve (purple) shows that when the storage was relatively inefficient, VDR exceeded VoS, which was reversed as the storage efficiency improved.

\section{Conclusion}\label{sec:conclusion}
This work analyzed the structural properties of the optimal policy co-optimizing flexible demands and storage devices when operated with a renewable DG. The policy is shown to abide by a load priority ranking rule that exercises consumption decisions based on load importance, which gets relaxed when the renewables are abundant. Comparative statics on the optimal decisions, prosumer payment, and reward have been investigated under different tariff and household DER parameters.  Lastly, it has been shown that under the special case of inflexible demands, the storage is operated in a manner that minimizes the inflows and outflows from and to the grid. 

{
\bibliographystyle{IEEEtran}
\bibliography{TPEC}
}

\newpage
\section*{Appendix: Proofs}\label{sec:appendix_proofs}
\subsection{Proof of proposition \ref{prop:loadranking}} 
From the solution of (\ref{eq:optimization}), the prosumer's fixed consumptions are functions of the prices $\pi^+_t, \pi^-_t, \gamma/\rho, \tau \gamma$ with the order in (\ref{eq:sandwichedGamma}). From the monotonicity of the marginal utilities $L_{tk}(\cdot)$ for every $t=0,\ldots,T-1$ and device $k$, the consumptions are ordered as: 
\beq \label{eq:ConsRank}
f_{tk}(\pi^-_t) \geq f_{tk}(\tau \gamma) \geq f_{tk}(\gamma/\rho) \geq f_{tk}(\pi^+_t)
\eeq
The proof of cases 1--5 in the proposition is in order:
\begin{enumerate}
    \item If $L_{tk}(0) > \pi^+_t$, then $L^{-1}_{tk}(\pi^+_t)>0$ hence $f_{tk}(\pi^+_t) >0$. Given (\ref{eq:ConsRank}) and the monotonicity of $d^\ast_{tk}$, device $k$'s minimum limit constraint will never bind. As a result, the device is consumed in all three net-consumption zones.
    \item If $\frac{\gamma}{\rho} < L_{tk}(0) \le \pi^+_t $, then $L^{-1}_{tk}(\pi^+_t)\leq 0$ and $L^{-1}_{tk}(\gamma/\rho)\geq 0$, hence $f_{tk}(\gamma/\rho)> 0$. Because $\underline{d}_k=0$, the consumption in the net-consumption zone is $f_{tk}(\pi^+_t)= 0$. Therefore, device $k$ is not consumed in the net-consumption zone, i.e. when $g_t\leq \Delta^+_t, \forall t$.
    \item If $\tau \gamma < L_{tk}(0)\le \frac{\gamma}{\rho}$, then $L^{-1}_{tk}(\gamma/\rho) \leq 0$ hence $f_{tk}(\gamma/\rho)= 0$. However, $L^{-1}_{tk}(\tau \gamma) > 0$ hence $f_{tk}(\tau \gamma)> 0$. Given (\ref{eq:ConsRank}) and the monotonicity of $d^\ast_{tk}$, the device $k$ is not consumed whenever $g_t \leq  \sigma^{+o}_t$ for every $t$.
    \item If $\pi^-_t < L_{tk}(0) \leq \tau \gamma $, then $L^{-1}_{tk}(\tau \gamma) \leq 0$ hence $f_{tk}(\tau \gamma)= 0$. However, $L^{-1}_{tk}(\pi^-_t) > 0$ hence $f_{tk}(\pi^-_t)> 0$. Given (\ref{eq:ConsRank}) and the monotonicity of $d^\ast_{tk}$, the device $k$ is not consumed whenever $g_t \leq  \sigma^{-}_t$.
    \item If $ L_{tk}(0) < \pi^-_t$, then $L^{-1}_{tk}(\pi^-_t)<0$ hence $f_{tk}(\pi^-_t)= 0$. Given (\ref{eq:ConsRank}) and the monotonicity of $d^\ast_{tk}$, the device $k$ is never consumed. \QED
\end{enumerate}

\subsection{Proof of proposition \ref{prop:ModulusMinimize}}
When the consumption is dropped from the set of decision variable, the MDP {\em policy} $\mu := (\mu_0,\ldots,\mu_{T-1})$ becomes a sequence of decision rules, $x_t \stackrel{\mu_t}{\rightarrow} u_t := e_t$, for all $x_t$ and $t$, that specifies storage operation in each interval. The optimal storage operation of the prosumer decision problem in (\ref{eq:optimization}), under A1-A2, follows from Proposition 3 in \cite{Alahmed&Tong:22arXiv}, with fixing the consumption at $f_{t}(\pi^+_t)$ for every $t=0,\ldots,T-1$, as
\beq 
e^\ast_t(g_t) = \left\{\begin{array}{ll}
\max\{g_t-f_{t}(\pi^+_t), -\underline{e}\} & g_t \le f_{t}(\pi^+_t)\\
\min\{g_t-f_{t}(\pi^+_t), \bar{e}\} & g_t> f_{t}(\pi^+_t).\\
\end{array}
\right.\nn
\eeq
Given
\bea
\max (a, b) \pm c &= \max (a \pm c, b \pm c )\nn\\
\min (a, b) \pm c &= \min (a \pm c, b \pm c),\nn
\eea
the optimal net-consumption $z^\ast_t(g_t):=f_{t}(\pi^+_t)-g_t+e^\ast_t$, can be written as:
\beq 
z^\ast_t(g_t) = \left\{\begin{array}{ll}
\max\{0, f_{t}(\pi^+_t)-g_t-\underline{e}\}, & z^\ast_t(g_t) \geq 0\\
\min\{0, f_{t}(\pi^+_t)-g_t+\bar{e}\}, & z^\ast_t(g_t) < 0,\\
\end{array}
\right.\nn
\eeq
which is simply
\beq \label{eq:PropOptNetCons}
z^\ast_t(g_t) = \left\{\begin{array}{ll}
f_{t}(\pi^+_t)-g_t-\underline{e}, & z^\ast_t(g_t) > 0\\
0, & z^\ast_t(g_t) =0\\
f_{t}(\pi^+_t)-g_t+\bar{e}, & z^\ast_t(g_t) < 0.\\
\end{array}
\right.
\eeq
Note that to solve (\ref{eq:PassiveOptimization}), for every $t=0,\ldots,T-1$, one can break:
\bea
\mathcal{P}: & \underset{e \in \{-\underline{e},\bar{e}\}}{\text{minimize}} & |z_t|,\nonumber
\eea
to the following three convex optimizations $\mathcal{P}^+, \mathcal{P}^-$ and $\mathcal{P}^0$:
\bea
\mathcal{P}^+: & \underset{e_t \in \{-\underline{e},\bar{e}\}}{\text{minimize}} & z_t \\
& \text{subject to} & f_{t}(\pi^+_t)-g_t + e_t \geq 0 \nonumber\\
\mathcal{P}^-: & \underset{e_t \in \{-\underline{e},\bar{e}\}}{\text{minimize}} & -z_t \\
& \text{subject to} & f_{t}(\pi^+_t)-g_t + e_t \leq 0 \nonumber\\
\mathcal{P}^0: & \underset{e_t \in \{-\underline{e},\bar{e}\}}{\text{minimize}} & z_t \\
& \text{subject to} & f_{t}(\pi^+_t)-g_t + e_t = 0 \nonumber
\eea

Given $g_t$, the optimal schedule is the one that achieves the minimum value among $\mathcal{P}^+, \mathcal{P}^-$ and $\mathcal{P}^0$. Note that, for all three optimizations, the optimal storage operation exist. Because the Slater’s condition is satisfied for these optimizations, KKT conditions for optimality is necessary and sufficient.
\par From Theorem 1 in \cite{Alahmed&Tong:22IEEETSG}, we can use the renewable adjusted consumption $\tilde{d}_t:= f_{t}(\pi^+_t)-g_t$ here to characterize the optimal storage operation, as:
\beq \label{eq:PropOptStorage}
e^\ast_t(g_t) = \left\{\begin{array}{ll}
-\underline{e}, & -\tilde{d}_t< -\underline{e} \\
-\tilde{d}_t, & -\tilde{d}_t \in [-\underline{e},\overline{e}]\\
\overline{e}, & -\tilde{d}_t> \overline{e}.\\
\end{array}
\right.
\eeq
Using the optimal storage operation $e^\ast_t(g_t)$ the optimal net consumption is
\bea \label{eq:PropOptNetConsOtherDirection}
z^\ast_t(g_t) &=& \tilde{d}_t+e^\ast_t(g_t)\nonumber\\
&=& \left\{\begin{array}{ll}
\tilde{d}_t-\underline{e}, & z^\ast_t(g_t)>0 \\
0, & z^\ast_t(g_t)=0\\
\tilde{d}_t+\overline{e}, & z^\ast_t(g_t)<0,
\end{array}
\right.\nonumber
\eea
which is equivalent to (\ref{eq:PropOptNetCons}). \QED

\subsection{Proof of corollary \ref{prop:ZeroZoneLengthComparison}} 
The proof of the corollary is based on the thresholds of the operational zones of every customer type. For every $t=0,\ldots,T-1$, the net-zero zone of active SDG prosumers is when $g_t\in [\Delta^+_t,\Delta^-_t]$, hence the length of the zone is 
\[|\Gc_t^{\bm{\mu}^\ast_4}| =\Delta^-_t-\Delta^+_t= f_{t}(\pi^-_t) - f_{t}(\pi^+_t) +\bar{e}+\underline{e}. \]
The net-zero zone of passive SDG prosumers with consumption $f_{t}(\pi^+_t)$ is when $g_t\in [f_{t}(\pi^+_t)-\underline{e},f_{t}(\pi^+_t)+\overline{e}]$ (as shown in the proof of proposition \ref{prop:ModulusMinimize}), hence the length of the zone is 
\[|\Gc_t^{\bm{\mu}^\ast_3}|= \bar{e}+\underline{e}. \]
The net-zero zone of active DG prosumers is when $g_t \in [f_{t}(\pi^+_t),f_{t}(\pi^-_t)]$ (as Theorem 1 in \cite{Alahmed&Tong:22IEEETSG} shows), hence the length of the zone is
\[|\Gc_t^{\bm{\mu}^\ast_2}| = f_{t}(\pi^-_t) -f_{t}(\pi^+_t). \]
 Lastly passive DG prosumers have the smallest net-zero zone length $|\Gc_t^{\bm{\mu}^\ast_1}|$ because $z^\ast$ crosses zero only in one distinct point, i.e., $f_{t}(\pi^+_t) = g_t$.
 
 Therefore, the ranking in (\ref{eq:NZranking1}) holds if $f_{t}(\pi^-_t)-f_{t}(\pi^+_t)\leq \bar{e}+\underline{e}$, and the ranking in (\ref{eq:NZranking2}) holds if $f_{t}(\pi^-_t)-f_{t}(\pi^+_t)> \bar{e}+\underline{e}$. One should easily verify (\ref{eq:NZranking3}) from the expressions of $|\Gc_t^{\bm{\mu}^\ast_2}|, |\Gc_t^{\bm{\mu}^\ast_3}|$, and $|\Gc_t^{\bm{\mu}^\ast_4}|$ above. \QED

\subsection{Theorem \ref{thm:ComparativeStatics} and its proof}  
 \begin{theorem}[Comparative statics analysis of NEM X prosumer]\label{thm:ComparativeStatics}
Under A1-A2, for every $t=0,\ldots,T-1$, and every device $k$, the optimal consumption level $d_{tk}^\ast(\cdot)$, optimal storage operation $e^\ast_t(\cdot)$, and optimal stage reward $S^{\ast,\pi_t}_t(\cdot)$ are all
monotonically increasing with $g_t$, whereas the prosumer payment $P^{\ast,\pi_t}_t(\cdot)$ is monotonically decreasing with $g_t$.

\par For every $g_t$ at the interior of each scheduling zone, 1) the consumption $d^\ast_{tk}$ monotonically decreases with $\pi^+_t, \pi^-_t$ and $\gamma$, 2) the storage operation is independent of $\pi^+_t$ and $\pi^-_t$, but it monotonically increases with $\gamma$ in the net-zero zone, 3) the payment $P^{\ast,\pi_t}_t(\cdot)$ monotonically decreases with $\pi^-_t$, and independent of $\gamma$, and lastly 4) the optimal surplus $S^{\ast,\pi_t}_t(\cdot)$ monotonically decreases with $\pi^+_t$ and $\gamma$ and monotonically increases with $\pi^-_t$ and $\gamma$.

\par The fixed charges $\pi^0$ do not affect $d_{tk}^\ast(\cdot)$ and $e^\ast_t(\cdot)$, but $S^{\ast,\pi_t}_t(\cdot)$ and $P^{\ast,\pi_t}_t(\cdot)$ monotonically decreases and monotonically increases with $\pi^0$, respectively.

\end{theorem}

\subsection*{Proof of Theorem \ref{thm:ComparativeStatics}}
We prove Theorem \ref{thm:ComparativeStatics} in three stages: 1) comparative statics of $d_{tk}^\ast$ and $e^\ast_t$, 2) comparative statics of $P^{\ast,\pi_t}_t$, and 3) comparative statics of $S^{\ast,\pi_t}_t$. \\
1) \underline{\textit{Comparative statics of $d_{tk}^\ast$ and $e^\ast_t$}}: Under A1-A2, and for every $t=0,\ldots,T-1$, and every device $k$, the storage operation $e^\ast_t$ and consumption $d_{tk}^\ast$ expressions under the optimal co-optimization policy are given by \cite{Alahmed&Tong:22arXiv}
\bea
e^\ast_t&=&\left\{\begin{array}{ll}
-\underline{e}, & g_t \le \sigma^+_t\\
g_t-\sigma_t^{+o}, & g_t\in [\sigma^+_t, \sigma^{+o}_t]\\
0, & g_t \in [\sigma^{+o}_t, \sigma^{-o}_t]\\
 g_t- \sigma_t^{-o}, & g_t \in [\sigma^{-o}_t, \sigma^-_t]\\
\overline{e}, &  g_t \ge \sigma^-_t,\\
\end{array}\right.\nonumber \\
d_{tk}^\ast &=& \left\{\begin{array}{ll}
f_{tk}(\pi^+_t), & g_t \le \Delta^+_t\\
f_{tk}(f^{-1}_t(g_t+\underline{e})), & g_t \in (\Delta_{t}^+,\sigma_{t}^+]\\
f_{tk}(f^{-1}_t(\gamma/\rho)), & g_t\in (\sigma^+_t, \sigma^{+o}_t)\\
f_{tk}(f^{-1}_t(g_t)), &  g_t \in [\sigma^{+o}_t, \sigma^{-o}_t]\\
f_{tk}(f^{-1}_t(\gamma\tau)), & g_t\in (\sigma^{-o}_t,\sigma_t^-)\\
f_{tk}(f^{-1}_t(g_t-\overline{e})), & g_t\in [\sigma^{-}_t,\Delta_t^-)\\
f_{tk}(\pi^-_t), & g_t \ge \Delta_{t}^-,\\
\end{array}\right. \nonumber
\eea
respectively. Theorem 1 in \cite{Alahmed&Tong:22arXiv} showed that the storage operation and consumption are both monotonically increasing functions of $g_t$.  Additionally, the monotonicity of $L_{tk}$ indicates that the optimal consumption $d_{tk}^\ast$ decreases with $\pi^+_t, \gamma$  and $\pi^-_t$. From the expression of $e^\ast_t$ above, the storage operation is independent of $\pi^+_t$ and $\pi^-_t$, but monotonically increases with $\gamma$ in the net-zero zone.

2) \underline{\textit{Comparative statics of $P^{\ast,\pi_t}_t$}}: Under A1-A2, and for every $t=0,\ldots,T-1$, the net consumption under the optimal co-optimization policy $z^\ast_t = d^\ast_t -g_t + e^\ast_t$ is a monotonically decreasing functions of $g_t$, as shown in Theorem 1 in \cite{Alahmed&Tong:22arXiv}.
Using the prosumer payment definition in (\ref{eq:NEMpayment}) and the optimal net consumption, the payment under the optimal co-optimization policy $P^{\ast,\pi_t}_t$ becomes:
\[P^{\ast,\pi_t}_t =\pi^0_t + \left\{\begin{array}{ll}
\pi^+(f_{t}(\pi^+_t) - g_t - \underline{e}),   &  g_t < \Delta^+_t\\
 0, &  g_t \in [\Delta^+_t, \Delta^-_t]\\
\pi^-(f_{t}(\pi^-_t) - g_t + \overline{e}),   &  g_t > \Delta^-_t,
\end{array}\right. \]
which is monotonically decreasing with $g_t$ and monotonically increasing with $\pi^0_t$. The payment is monotonically decreasing with $\pi^-_t$, as $\frac{\partial f_{t}(\pi^-_t)}{\partial \pi^-_t}\leq 0$. The payment is independent of $\gamma$, as $\frac{\partial P^{\ast,\pi_t}_t}{\partial \gamma}=0$.

3) \underline{\textit{Comparative statics of $S^{\ast,\pi_t}_t$}}: Under A1-A2, and for every $t=0,\ldots,T-1$, and every device $k$, the monotonicity of $S^{\ast,\pi_t}_t$ is shown by recalling that, because $U_{tk}$ is monotonically increasing with $d^\ast_{tk}$, which is monotonically increasing with $g_t$, it holds that $U_{tk}(g_t)$ is a monotonically increasing function of $g_t$. Given that $U_{tk}$ and $P^{\ast,\pi_t}_t$ are monotonically increasing and monotonically decreasing with $g_t$, respectively, the prosumer surplus is monotonically increasing with $g_t$.
\par Increasing $\pi^+$ monotonically decreases $S^{\ast,\pi_t}_t$, because in the net consumption zone ($g_t<\Delta^+_t$), we have
\[ S^{\ast,\pi_t}_t=\sum_{k=1}^K U_{tk}(f_{tk}(\pi^+_t))-\pi^+(\sum_{k=1}^K f_{tk}(\pi^+_t) - g_t - \underline{e})-\pi^0_t,\]
and deriving the above expression with respect to $\pi^+_t$, gives $-(\sum_{k=1}^K f_{tk}(\pi^+_t) - g_t - \underline{e})<0$.
\par Increasing $\pi^-$ monotonically increases $S^{\ast,\pi_t}_t$, because in the net production zone ($g_t>\Delta^-_t$), we have
\[ S^{\ast,\pi_t}_t=\sum_{k=1}^K U_{tk}(f_{tk}(\pi^-_t))-\pi^-(\sum_{k=1}^K f_{tk}(\pi^-_t) - g_t + \overline{e})-\pi^0_t,\]
and deriving the above expression with respect to $\pi^-_t$, gives $-(\sum_{k=1}^K f_{tk}(\pi^-_t) - g_t - \overline{e})>0$.
\par Increasing $\gamma$ monotonically decreases $S^{\ast,\pi_t}_t$, because in the net zero zone ($g_t\in [\Delta^+_t,\Delta^-_t]$), the surplus is simply $S^{\ast,\pi_t}_t=\sum_{k=1}^K U_{tk}(d^\ast_{tk})$, and given that $d^\ast_{tk}$ monotonically decreases with $\gamma$, it holds that the surplus is also a monotonically decreasing function of $\gamma$. 
\par Finally, the monotonicity of $\pi^0_t$ is immediate from the definitions.
\QED

\end{document}